\documentclass[10pt]{revtex4}
\usepackage{amssymb,amsmath,amscd}
\usepackage{graphicx,calc,epsfig,pstricks,bbm}
\usepackage{tikz}

\expandafter\ifx\csname package@font\endcsname\relax\else
 \expandafter\expandafter
 \expandafter\usepackage
 \expandafter\expandafter
 \expandafter{\csname package@font\endcsname}%
\fi

\newcommand {\be}{\begin{equation}}
\newcommand {\ee}{\end{equation}}
\newcommand{\ba}{\begin{array}{c}}
\newcommand{\ea}{\end{array}}

\ifx\JPicScale\undefined\def\JPicScale{1}\fi
\unitlength \JPicScale mm

\begin{document}
\title{Physical states in Quantum Einstein-Cartan Gravity}%

\author{Francesco Cianfrani}%
\email{francesco.cianfrani@ift.uni.wroc.pl}
\affiliation{Institute
for Theoretical Physics, University of Wroc\l{}aw, Pl.\ Maksa Borna
9, Pl--50-204 Wroc\l{}aw, Poland.}
\date{\today}%

\begin{abstract} 
The definition of physical states is the main technical issue of canonical approaches towards Quantum Gravity. In this work, we outline how those states can be found in Einstein-Cartan theory via a continuum limit and they are given by finite dimensional representations of the Lorentz group. 
\end{abstract}


\maketitle

\section{Introduction}

The quantization of General Relativity (GR) has interpretative and technical issues, which up to now prevented to successfully achieve a widely accepted quantum theory and suggested to look for alternatives, the most prominent being String Theory. In this work, we want to support the idea that canonical quantization can be performed, by proposing a formulation in which one can find physical states for GR in vacuum.    

Physical states are the basic objects in the canonical quantization based on the Dirac prescription, in which the classical constraints are promoted to operators annihilating them in a suitable space. In geometrodynamics, to find physical states essentially means to solve the Wheeler-de Witt (WdW) equation \cite{DeWitt:1967yk}, which has been done only in symmetry reduced models \cite{Misner:1969ae}, the full theory being elusive. In Loop Quantum Gravity (LQG), several achievements concerning the definition of the operators associated with the constraints have been done \cite{Ashtekar:1995zh}. These have been obtained via a reformulation of GR in terms of some $SU(2)$ connections \cite{Ashtekar:1986yd,Barbero:1994ap} and the use of some tools proper of lattice gauge theories \cite{Wilson:1974sk,Kogut:1974ag}, such as the description of quantum states in terms of holonomies. A result of LQG is that one knows in which space to look for physical states, but, due to the complicated expression of the operators in terms of basic variables, one cannot get any explicit expression for them \cite{Thiemann:1996aw}. 

One can interpret these difficulties as due to the peculiar structure of GR with respect to the other interactions for which the quantization tools have been developed. A peculiarity of GR is that the action contains second derivatives of fields. Indeed, a formulation with up to first derivatives exists, Einstein-Cartan gravity. We are going to outline how the constraints in Einstein-Cartan gravity take a simple form in loop representation, which allows us to manage them as quantum operators acting on Wilson loops of the spin connections. The issue is now that the system of constraints is second-class and the Dirac prescription is generically inconsistent. We propose a way to circumvent the second class character of the system of constraints on a quantum level, by performing a continuum limit. 

The continuum limit is a key-point of any quantum theory of gravity. In WdW formulation, one deals with a continuous model, in which fields operators act as distributions. This poses several problems concerning regularization, since infinities are known to arise (see for instance \cite{KowalskiGlikman:1996ad,Blaut:1997zr}). A different point of view is commonly accepted in LQG. The idea is that the final theory is discrete and finite, such that once a discretization has been introduced, all physical quantities (included constraints) are finite. Hence, the continuum limit is just a tool to infer the low-energy limit of the theory, but the theory is fundamentally discrete. This interpretation is supported by the general expectation of a discrete geometry in Quantum Gravity.   

Here instead, we consider a model which makes sense only in the continuum and we use the discretization only as a tool to solve the model. We will see how after the removal of the regulator ($\epsilon\rightarrow 0$) some operator diverges. This is not necessary an issue, since they are distributions ab-initio, and we require the corresponding constraints to hold as distributions as well, {\it i.e.} modulo some diverging ($1/\epsilon$) factors. This is the same kind of interpretation adopted in lattice gauge theories, in which the lattice is not physical. 

In particular, we will consider as states the Wilson loops of the spin connection, which are constructed out of holonomies of the Lorentz group. This will lead us to consider irreducible representations of the Lorentz group, which are generically infinite-dimensional. 
The study of irreducible representations of the Lorentz group has been originally performed in connection with conformal covariant models of interacting scalar fields (see \cite{Dobrev:1976vr} and references therein). Then, it has been reconsidered in Quantum Gravity in order to give a Lorentz covariant description of the $SU(2)$ connections of LQG \cite{Alexandrov:2001wt,Alexandrov:2002br,Engle:2007wy,Alexandrov:2010un,Ding:2010fw}. However, the description in terms of Lorentz representations has been proposed only as an intermediate step, the final aim being getting $SU(2)$ representations. In fact, on a quantum level a crucial technical simplification occurs when working with a compact group: the measure can be inherited from the Haar measure of the group which is bounded. 

Here, we consider ab-initio Lorentz holonomies and decompose into irreducible Lorentz representation, so we deal with un-normalized states already on a kinematical level, but we will outline how the expression of the scalar constraint simplifies with respect to LQG. The result of this analysis is a condition fixing the quantum numbers of the Wilson loops which are annihilated by the scalar constraint. We will demonstrate that a solution exists and it is given by finite dimensional representations. 

\section{Hamiltonian analysis of Einstein-Cartan gravity} 

The analysis of the Einstein-Cartan action in the tetrad formalism is relatively simple and shares some similarities with topological BF theory \cite{Birmingham:1991ty} (see \cite{Freidel:2012np} for a comparison). The action can be written as (in units $c=8\pi G=1$)
\be
S=\frac{1}{2}\int e\,e^\mu_I\,e^\nu_J\,R^{IJ}_{\mu\nu}\,d^4x\,,\label{EH}
\ee
$e^\mu_I$ being inverse tetrads, which are related to the metric tensor $g_{\mu\nu}=\eta_{IJ}\,e^I_\mu\,e^J_\nu$,
$e$ is the determinant of $e^I_\mu$, while $R^{IJ}_{\mu\nu}$ is the curvature of the spin connection $\omega^{IJ}_\mu$, namely
\be
R^{IJ}_{\mu\nu}=\partial_\mu\omega^{IJ}_\nu-\partial_\nu\omega^{IJ}_\mu+\omega^{IK}_\mu\,\omega^{\phantom1J}_{K\phantom1\nu}-\omega^{IK}_\nu\,\omega^{\phantom1J}_{K\phantom1\mu}\,.
\ee
In BF theory the Hodge dual of a generic two form $B^{IJ}_{\mu\nu}$ replaces $e\,e^\mu_I\,e^\nu_J$ and the theory is topological since from the variation with respect to $B^{IJ}_{\mu\nu}$ it follows that the curvature vanishes. 
  
The total Hamiltonian is a linear combination of the following constraints \cite{Cianfrani:2008zv} (see also \cite{libro})

\begin{equation}   
\begin{split}
&\mathcal{S}=\pi^a_{IK}\,\pi^{bK}_{\phantom1J}\,R^{IJ}_{ab}=0 \\\\ 
&\mathcal{V}_a= \pi^b_{IJ}\,R^{IJ}_{ab}=0 \\\\ 
&G_{IJ}=D^{(\omega)}_a\pi^{a}_{IJ}=\partial_a\pi^a_{IJ}-2\omega_{a[I}^{\phantom1\phantom2K}\,\pi^a_{|K|J]}=0 \\\\ 
&C^{ab}=\epsilon^{IJKL}\,\pi_{IJ}^{(a}\,\pi_{KL}^{b)}=0 \\\\ 
&D^{ab}=\epsilon^{IJKL}\,\pi^c_{IM}\,\pi^{(aM}_{\phantom1\phantom2J}\,D^{(\omega)}_c\pi^{b)}_{KL}=0
\end{split}.\label{hcon}
\end{equation} 
where $\pi^a_{IJ}$ denotes the conjugate momentum of $\omega^{IJ}_a$. The first two constraints are the scalar and vector constraints, which modulo the third one, coincide with the supermomentum and superHamiltonian constraints of the ADM formulation. Hence, they are associated with the invariance under coordinate transformations in the 3+1 representation. The third constraint is the Gauss constraint generating local Lorentz transformations and it arises in view of the invariance of the whole formulation under Lorentz transformations acting on internal indexes (which by definition do not change the form of the metric). 
 
The last two constraints make the whole system of constraints second-class, since $\{C^{ab},D^{cd}\}$ and $\{D^{ab},D^{cd}\}$ do not vanish on the constraint hypersurface.

On a quantum level, the role of second-class constraints is more elusive than in classical physics. The standard Dirac procedure for quantization of first class constraints cannot generically be applied. In fact, let us suppose to have two second-class constraints $A_1=0$ and $A_2=0$, with $\{A_1,A_2\}=B\neq0$. If we quantize the system and we consider same states $\psi$ in the kernel of the operator $\hat{A}_1$, {\it i.e.} $\hat{A}_1\psi=0$, then the action of the operator $\hat{A}_2$ takes $\psi$ out of the kernel, since 
\be
\hat{A}_1\,\hat{A}_2 \psi = \hat{A}_2\,\hat{A}_1 \psi + \hat{B} \psi= \hat{B} \psi\,.
\ee
This is the case, unless one can solve both $\hat{A}_1\psi=0$ and $\hat{A}_2\psi=0$, which imply that the operator associated to $B$ vanishes. Of course, one should verify a posteriori whether such a restriction of states is too strong.  

Returning to gravity, it is worth noting how the constraint $D^{ab}$ makes the whole system of constraints second-class, since the non vanishing Poisson brackets are $\{C^{ab},D^{cd}\}$ and $\{D^{ab},D^{cd}\}$. A closer inspection on how the constraints \eqref{hcon} emerge reveals that 
\be
D^{ab}= \{C^{ab},\mathcal{S}\}+\ldots\,,\label{D}
\ee
where $\ldots$ denote terms proportional to the Gauss constraint. 
In what follows, we will show that it is possible to define some states $\psi$ which are in the kernel of the operator corresponding to $C^{ab}$ and that we can construct $\hat{\mathcal{S}}$ such that its action does not take states out of the kernel in the continuum limit. This achievement solves on a quantum level all the issues related with the second class character of the system of constraints \eqref{hcon}, since the operator associated to $D^{ab}$, defined from \eqref{D}, vanishes. Moreover, we will also derive a class of states which are annihilated by $\mathcal{S}$. These states are naturally candidates for being physical states of quantum gravity.

\section{Wilson loops of the Lorentz group} 

Let us consider a loop representation for quantum gravity, in which states are defined as Wilson loops $W_\alpha(\omega)$ of the spin connection along some loop $\alpha$, {\it i.e.} as the trace of the holonomy along $\alpha$
\be
W^{k,\rho}_{\alpha}(\omega)=Tr\left[h^{k,\rho}_\alpha(\omega)\right]=Tr\Bigg[P\left(\exp\int_\alpha \omega^{IJ}_a dx^a\,\tau^{k,\rho}_{IJ}\right)\Bigg]\,,
\ee 
$\tau^{k,\rho}_{IJ}$ being Lorentz generators in the irreducible representation $(k,\rho)$. Here, we refer to Naimark classification of Lorentz irreducible representations \cite{Naimark} (a brief introduction is also given in \cite{Cianfrani:2010jf}), in which $k$ is a integer or semi-integer non-negative number and $\rho\in\mathbb{C}$. Each representation is constructed as a tower of $SU(2)$ irreducible representations with spin numbers $j$ from $k$ up to infinity, {\it i.e.}
\be
W^{k,\rho}=\oplus_{j=k}^{+\infty}\, W^{j}\,. 
\ee
Let us stress that working with holonomies of the Lorentz group is much more complicated than, for instance, with $SU(2)$ holonomies. The Haar measure of Lorentz group is unbounded, since the group is noncompact, and irreducible representations are generically infinite-dimensional. Hence, we cannot rigorously define the kinematical Hilbert space as in LQG.

There are two different kinds of representations: the complementary series for $k=0$ and imaginary $\rho$ (and it is unitary for $0\leq |\rho| \leq 1$), the principal series for real $\rho$. 

Momenta $\pi^a_{IJ}$ are smeared over surfaces $S$ and their action provides the insertion of a Lorentz generator at the intersection point $P$ between $\alpha$ and $S$ (if there is no intersection the momentum operator vanishes)
\be
\hat{\pi}_{IJ}(S)\,W_\alpha(\omega)=i\,o_{(\alpha,S)}\, Tr\left[h_{\alpha_1}\,\tau_{IJ}\,h_{\alpha_2}\right]\,,\label{pi}
\ee
where $o_{(\alpha,S)}=\pm1,0$ depending on the relative orientation between $S$ and $\alpha$, and $\alpha=\alpha_1\cup\alpha_2\quad P=\alpha_1\cap\alpha_2$. 
These states are invariant under local Lorentz transformations, thus the condition $\hat{G}_{IJ}\,W_{\alpha}=0$ holds. The vector constraint $\mathcal{V}_a$ generates spatial diffeomorphisms and, mimicking the procedure adopted in Loop Quantum Gravity, one can formally solve it by defining states over s-knots $[\alpha]$, {\it i.e.} over the equivalence classes of loops under diffeomorphisms \cite{Ashtekar:1995zh}.    

The operators associated to $C^{ab}(x)$ can be defined via a proper regularization prescription. We can replace the two momenta with the associated fluxes across two surfaces with infinitesimal area $\epsilon^2$ \footnote{There is no background structure, thus no metric, but there are coordinates, which allow us to define coordinated length, area and volume}. The resulting expression converges to the continuum one in the limit $\epsilon\rightarrow0$ and it is ready to be quantized. This way, we get the following operator associated to $C^{ab}$
\be
\hat{C}^{ab}(x)=\lim_{\epsilon\rightarrow 0} \frac{1}{\epsilon^4} \,\epsilon^{IJKL}\,\hat{\pi}_{IJ}(S_x^a)\,\hat{\pi}_{KL}(S_x^b)\,,\label{10}
\ee
where $S_x^a$ denotes a surface containing $x$ dual to the direction $a$. It is worth noting that the operator above is not regularized, as it diverges in the limit $\epsilon\rightarrow 0$. This is not surprising, since its classical counterpart is quadratic in momenta, thus it is a distribution with the same degree of divergence. This also means that if the operator vanishes it does as a distribution (times a diverging factor in the continuum limit). Hence, one can define a renormalized operator $\hat{C}^{ab}_{ren}=\epsilon^{4}\,C^{ab}$ and work with only finite quantities. 

This way, one gets (see appendix \ref{A})
\be
\hat{C}^{ab}_{ren}(x)\,W^{k,\rho}_\alpha(\omega)=o_{(\alpha,S_x^a)}\,o_{(\alpha,S_x^b)}\,\,2k\rho\,W^{k,\rho}_\alpha(\omega)=0\,.\label{11}
\ee
and we have two possible solutions, $k=0$ and $\rho=0$. In what follows, we consider 
\be
k=0\,.\label{k0}
\ee

\section{Scalar Constraint} 

In order to construct the operator associated with the scalar constraint $\mathcal{S}$, we need to replace the curvature $R^{IJ}_{ab}$ with the corresponding expression in terms of holonomies $h^{k_0,\rho_0}_l$ along same link $l$. This is generically ambiguous, since the quantum numbers $\{k_0,\rho_0\}$ of the holonomy are not specified. The action of the holonomy operator on a Wilson loop provides the insertion of the holonomy into the quantum state, if the holonomy and the original loop do not share any link, while for common links the resulting representation is obtained via recoupling theory. The recoupling theory gives the rule to combine two representations into a new one, the most celebrated sample being $SU(2)$ recoupling theory, which tell us how to sum up spins via Clebsch-Gordan coefficients (or Wigner $3j$ symbols). Similarly, one can define Clebsch-Gordan coefficients for the Lorentz group \cite{Anderson:1970ez} and the product between two holonomies along a common link $l$ provides a linear combination of holonomies along the same link with Clebsch-Gordan coefficients. 

We want now to define the operator $\hat{\mathcal{S}}$ such that in the continuum limit it does not take us out of the space in which \eqref{k0} holds. In other words, we see the discretization of the spatial manifold as a pure regularization {\it i.e.} as a mere formal way to do computations, and we check the consistency of our results (in this case the consistency between the constraints $C^{ab}=0$ and $\mathcal{S}=0$) only after having removed the regulator. 

Let us now construct explicitly the operator $\hat{\mathcal{S}}(x)$, corresponding to $\mathcal{S}(x)$. Similarly to the operator $\hat{C}^{ab}$, $\mathcal{S}$ can be defined as follows
\begin{equation}
\hat{\mathcal{S}}(x)=\lim_{\epsilon\rightarrow 0} \frac{A}{\epsilon^6}\,\hat{\mathcal{S}}_{ren}(x)\qquad 
\hat{\mathcal{S}}_{ren}(x)=\hat{\pi}^{I}_{\phantom1K}(S_x^a)\,\hat{\pi}^{KJ}(S_x^b)\,\, Tr\left[\tau^{0,\rho_0}_{IJ}\,h^{0,\rho_0}_{\Omega^x_{ab}}
\right]+Tr\left[\tau^{0,\rho_0}_{IJ}\,h^{0,\rho_0}_{\Omega^x_{ab}}\right
]\,\,\hat{\pi}^{KJ}(S_x^b)\,\hat{\pi}^{I}_{\phantom1K}(S_x^a)\,,\label{Sv}
\end{equation}
$\Omega^x_{ab}$ being a loop constructed with the dual links to $S^x_a$ and $S^x_b$, while $A$ denotes a normalization factor such that $Tr[\tau^{0,\rho_0}_{KL}\tau^{0,\rho_0}_{KL}]=A^{-1}\,\eta_{I[K}\, \eta_{L]J}$. In the expression above we choose a symmetric ordering between the holonomy and flux operators.
We introduced the renormalized operator $\hat{\mathcal{S}}_{ren}(x)$, in which we removed the $1/\epsilon^6$ factor and also the constant $A$, which diverges for infinite representations. 

The action of $\hat{\mathcal{S}}$ on $W_{\alpha}$ (see appendix \ref{B}) provides the insertion of the loop $\Omega^x_{ab}$ and of three Lorentz generators, corresponding to the two momenta and the Lorentz generator within the trace in \eqref{Sv}. The nontrivial case is that in which the loop $\Omega^x_{ab}$ and $\alpha$ have one edge in common, such that Lorentz recoupling theory has to be applied to infer the representation based at that edge. Even if we choose the quantum numbers $(k_0,\rho_0)$ of the holonomy along $\Omega^x_{ab}$ such that \eqref{11} holds, namely $k_0=0$, the recoupling theory between states with quantum numbers $(0,\rho)$ and $(0,\rho_0)$ provides some representations $(k',\rho')$ with $k'\neq 0$, such that resulting state is not anymore a solution of \eqref{11}. 
However, in the continuum limit the loop $\Omega^x_{ab}$ shrinks to a point and, taking care of the presence of Lorentz generators, one can see how the resulting state does not contain any representation violating the condition \eqref{k0}. In this sense, we circumvent the second-class character of the system of constraints \eqref{hcon}.

Furthermore, one can demonstrate how the condition $\hat{\mathcal{S}}_{ren}(x) W^{0,\rho}_\alpha(\omega)=0$ holds if
\be
\sum_{k'}\int d\mu(k',\rho')\,(-k'^2+\rho')^2= \rho^2\,\sum_{k'}\int d\mu(k',\rho') \,,\label{fc2}
\ee
where the sum and the integrals extends over all the admissible values of $k'$ and $\rho'$, {\it i.e.} those for which the Clebsch-Gordan coefficient of the Lorentz group $C^{0\rho j m}_{k'\rho' j' m'\,0\rho_0j_0m_0}$ is nonvanishing. Generically, it is very hard to manage an expression like that, because the integral extends over an unbounded domain. In what follows, we will outline how \eqref{fc2} is solved for finite dimensional representations. 

\section{Finite dimensional representations} 

Finite dimensional representations of the Lorentz group can be obtained as the direct product of two complexified $SU(2)$ representations $D^{j0}$ and $D^{0j'}$, which are associated with the two $SU(2)$ subalgebras contained into the Lorentz algebra. We can represent them as follows \cite{Rao}
\be
D^{jj'}=D^{j0} \times D^{0j'}\,,\label{djj'}
\ee
$D^{j0}$ and $D^{0j'}$ being $SU(2)$ representations analytically continued to complex angles and related via complex conjugation
\be
\left(D^{0j}\right)^*=D^{j0}\,. 
\ee

Finite dimensional representations in Naimark classifications have the following quantum numbers
\be
k=|j-j'|\qquad |\rho|=1+j+j'\,,
\ee
where $\rho$ is imaginary, and they describe a tower of $SU(2)$ irreducible representations with spin number from $k$ up to $j+j'$.
The recoupling theory can be inferred from that of $SU(2)$, as follows
\begin{align}
D^{j_1j'_1} \otimes D^{j_2j_2'}=& \left(D^{j_10}\otimes D^{j_20}\right) \otimes \left(D^{0j_1'}\otimes D^{0j_2'}\right) = \nonumber\\
&\oplus^{j_1+j_2}_{j=|j_1-j_2|} \oplus^{j'_1+j'_2}_{j'=|j'_1-j'_2|} D^{jj'}\,.\label{rec}
\end{align}

The same relations \eqref{djj'} and \eqref{rec} hold for Euclidean gravity, {\it i.e.} if the Lorentz group is replaced by $SO(4)$. Hence, all the conclusions we will infer for finite representations of the Lorentz group are valid in the Euclidean case too, but in that case the representation \eqref{djj'} describes a couple of independent $SU(2)$ representations with spins $j$ and $j'$.

Going back to gravity, \eqref{11} is solved for  
\be
j=j'\,,
\ee
such that we are interested in those states of the form $D^{jj}$. Taking $D^{j_0j_0}$ for the holonomy in the regularized expression of the scalar constraint, the admissible quantum numbers $k'$ and $\rho'$ can be directly computed from the recoupling theory \eqref{rec}, {\it i.e.}
\be
D^{j_0j_0}\otimes D^{jj} =
\oplus_{I=j-j_0}^{j+j_0}\oplus_{I'=j-j_0}^{j+j_0}D^{II'}\,.
\ee

Hence, there are $(2j_0+1)^2$ admissible values for $(k',\rho')$, corresponding to $(|I-I'|,1+I+I')$ for $I,I'=j-j_0,\ldots,j+j_0$, such that the condition \eqref{fc2} becomes a finite summation over them. In particular, the right-hand side of \eqref{fc2} becomes
\be
\rho^2\,\sum_{(k',\rho')} =(2j_0+1)^2 (2j+1)^2\,,\label{rhs}
\ee
where we used $\rho=2j+1$, while the left-hand side rewrites
\be
\sum_{(k',\rho')} ((\rho')^2-k'^2)= \sum_{I,I'=j-j_0}^{j+j_0} \left[(1+I+I')^2-(I-I')^2\right]\,.
\ee
Surprisingly, the explicit computation of the expression above gives
\be
\sum_{I,I'=j-j_0}^{j+j_0} \left[(1+I+I')^2-(I-I')^2\right]=(2j_0+1)^2 (2j+1)^2\,,
\ee
which coincides with \eqref{rhs}.

Therefore, the condition \eqref{fc2} holds identically for finite representations. This means that $D^{jj}$ are simultaneous solutions of both the condition $C^{ab}=0$ and the scalar constraint in the continuum limit. 

\section{Conclusions} 

We have derived a procedure to define a class of states which are annihilated by the constraints of gravity in Einstein-Cartan formulation. In particular, we avoided the issues concerning the second-class character of the system of constraints \eqref{hcon}, by defining the scalar constraint mapping states in the kernel of $\hat{C}$ within themselves. In order to do that, we need to perform a continuum limit first. This means that the theory is consistent only after the continuum limit has been taken. Hence, the adopted regularization is just a formal tool to perform some calculations and has nothing to do with a physical quantum description for gravity. It is worth noting how this is exactly the status of lattice-regularization in Quantum Field Theories, while in Quantum Gravity the general expectation of a fundamental discrete geometric structure suggested to promote the lattice to a real description of the physical reality.

It is worth noting the differences with LQG and Spin Foam models. We considered pure Einstein-Cartan action, thus no Immirzi parameter is present (although it can be included). We did not really define a kinematical Hilbert space on which the constraints are well-defined operators, we just require the preHilbert space to be large enough to contain holonomies and left/right-invariant vector fields. The kinematical scalar product is problematic (but not the physical one) because we are dealing with a non-compact gauge group and the associated Haar measure is unbounded. Furthermore, the constraints operators are not regularized, but contain a diverging factor in the continuum limit. In some sense, we are less ambitious and rigorous than LQG and, inspired by topological theories, we are just looking for some states which are solutions of constraints. This is legitimate, since the kinematical Hilbert space may not be necessary at all and, even if it can be defined, it does not necessarily give us information on the space of physical states (see for instance \cite{Dittrich:2007th}).

In the covariant version of LQG, one identifies the kinematical states (which are boundary states for Spin Foam models) as certain $SU(2)$ subrepresentations within each Lorentz irreps \cite{Ding:2010fw}. Here, one finds that physical states are finite dimensional Lorentz irreps, thus a finite tower of $SU(2)$ representations, and the action of boosts can be naturally implemented on them (while in LQG their action is trivial, being proportional to that of rotations). Hence, the physical states of our model are different from the kinematical states of covariant LQG. It cannot be excluded that the two kinds of states can be related via a sort of gauge-fixing of boosts, as soon as the present analysis is repeated with the Holst modification of Einstein-Cartan action. However, it is worth noting that our states are physical and they can be found just because of the simplifications occurring in the scalar constraint when local boosts are not fixed.

In order to gain some intuition whether the states we found can capture the relevant features of the gravitational field we need to perform a proper semiclassical limit and study the behavior of observables. The lack of observables and of proper semiclassical techniques requires a theoretical effort to try to characterize the behavior of our solutions. Up to now, our states just stand as mere candidates for a full quantum theory of gravity.   \\\\\\

{\bf Acknowledgment-}   
FC is supported by funds provided by the National Science Center under the agreement
DEC-2011/02/A/ST2/00294. 


\begin{thebibliography}{19}%

\bibitem{DeWitt:1967yk} 
  B.~S.~DeWitt,
  Phys.\ Rev.\  {\bf 160}, 1113 (1967).

\bibitem{Misner:1969ae} 
  C.~W.~Misner,
  Phys.\ Rev.\  {\bf 186}, 1319 (1969).

\bibitem{Ashtekar:1995zh} 
  A.~Ashtekar, J.~Lewandowski, D.~Marolf, J.~Mourao and T.~Thiemann,
  J.\ Math.\ Phys.\  {\bf 36}, 6456 (1995)
  doi:10.1063/1.531252
  [gr-qc/9504018].

\bibitem{Ashtekar:1986yd} 
  A.~Ashtekar,
  Phys.\ Rev.\ Lett.\  {\bf 57}, 2244 (1986).

\bibitem{Barbero:1994ap} 
  J.~F.~Barbero G.,
  Phys.\ Rev.\ D {\bf 51}, 5507 (1995)

\bibitem{Wilson:1974sk} 
  K.~G.~Wilson,
  Phys.\ Rev.\ D {\bf 10}, 2445 (1974).

\bibitem{Kogut:1974ag} 
  J.~B.~Kogut and L.~Susskind,
  Phys.\ Rev.\ D {\bf 11}, 395 (1975).

\bibitem{Thiemann:1996aw}
  T.~Thiemann,
  Class.\ Quant.\ Grav.\  {\bf 15} (1998) 839
  doi:10.1088/0264-9381/15/4/011
  [gr-qc/9606089].

\bibitem{KowalskiGlikman:1996ad} 
  J.~Kowalski-Glikman and K.~A.~Meissner,
  Phys.\ Lett.\ B {\bf 376}, 48 (1996)
  doi:10.1016/0370-2693(96)00268-7
  [hep-th/9601062].

\bibitem{Blaut:1997zr} 
  A.~Blaut and J.~Kowalski-Glikman,
  Phys.\ Lett.\ B {\bf 406}, 33 (1997)
  doi:10.1016/S0370-2693(97)00665-5
  [gr-qc/9706076].

\bibitem{Dobrev:1976vr} 
  V.~K.~Dobrev, G.~Mack, I.~T.~Todorov, V.~B.~Petkova and S.~G.~Petrova,
  Rept.\ Math.\ Phys.\  {\bf 9}, 219 (1976).

\bibitem{Alexandrov:2002br} 
  S.~Alexandrov and E.~R.~Livine,
  Phys.\ Rev.\ D {\bf 67}, 044009 (2003)
  doi:10.1103/PhysRevD.67.044009
  [gr-qc/0209105].

\bibitem{Alexandrov:2001wt} 
  S.~Alexandrov,
  Phys.\ Rev.\ D {\bf 65}, 024011 (2002)
  doi:10.1103/PhysRevD.65.024011
  [gr-qc/0107071].
	
\bibitem{Engle:2007wy} 
  J.~Engle, E.~Livine, R.~Pereira and C.~Rovelli,
  Nucl.\ Phys.\ B {\bf 799}, 136 (2008)	
	
\bibitem{Alexandrov:2010un} 
  S.~Alexandrov and P.~Roche,
  Phys.\ Rept.\  {\bf 506}, 41 (2011)
  doi:10.1016/j.physrep.2011.05.002
  [arXiv:1009.4475 [gr-qc]].	

\bibitem{Ding:2010fw} 
  Y.~Ding, M.~Han and C.~Rovelli,
  Phys.\ Rev.\ D {\bf 83}, 124020 (2011)

\bibitem{Birmingham:1991ty} 
  D.~Birmingham, M.~Blau, M.~Rakowski and G.~Thompson,
  Phys.\ Rept.\  {\bf 209}, 129 (1991).

\bibitem{Freidel:2012np} 
  L.~Freidel and S.~Speziale,
  SIGMA {\bf 8}, 032 (2012)
	
\bibitem{Cianfrani:2008zv} 
  F.~Cianfrani and G.~Montani,
  Phys.\ Rev.\ Lett.\  {\bf 102}, 091301 (2009)
  doi:10.1103/PhysRevLett.102.091301
  [arXiv:0811.1916 [gr-qc]].

\bibitem{libro} 
  F.~Cianfrani, O.~M.~Lecian, M.~Lulli and G.~Montani,
  `` Canonical Quantum Gravity,''
  (World Scientific, 2014).

\bibitem{HT}
  M.~Hennaux and C.~Teitelboim,
  ``Quantization of Gauge Systems,''
  (Princeton University Press, 1994).
						

\bibitem{Naimark}			
M.~A.~Naimark, 
``Linear Representations of the Lorentz Group,'' 
(Pergamon Press, 1964).	
			
\bibitem{Cianfrani:2010jf} 
  F.~Cianfrani,
  Class.\ Quant.\ Grav.\  {\bf 28}, 175014 (2011)
  doi:10.1088/0264-9381/28/17/175014
  [arXiv:1012.1982 [gr-qc]].						
		
						
\bibitem{Anderson:1970ez} 
  R.~L.~Anderson, R.~Raczka, M.~A.~Rashid and P.~Winternitz,
  J.\ Math.\ Phys.\  {\bf 11}, 1050 (1970).

\bibitem{Anderson:1967zz} 
  R.~L.~Anderson, R.~Raczka, M.~A.~Rashid and P.~Winternitz,
  IC-67-50.
	url: http://streaming.ictp.trieste.it/preprints/P/67/050.pdf

\bibitem{Rao}			
K.~N.~Srinivasa~Rao, 
``The rotation and Lorentz groups and their representations for physicists,'' 
(Wiley, 1988).

						
\bibitem{Dittrich:2007th} 
  B.~Dittrich and T.~Thiemann,
  J.\ Math.\ Phys.\  {\bf 50}, 012503 (2009)
  doi:10.1063/1.3054277
  [arXiv:0708.1721 [gr-qc]].
						
\end{thebibliography}


%

\appendix

\section{Action of the constraint $\hat{C}^{ab}$}\label{A}

In this section, we compute explicitly the action of the constraint \eqref{10}. By using \eqref{pi} we get
\begin{equation}
\hat{C}^{ab}(x)\,W^{k,\rho}_\alpha(\omega)=\lim_{\epsilon\rightarrow 0}\,\frac{1}{\epsilon^4}  \hat{C}_{ren}^{ab}(x)\,W^{k,\rho}_\alpha(\omega)=
-\,o_{(\alpha,S_x^a)}\,o_{(\alpha,S_x^b)}\,\frac{1}{\epsilon^4} \,\epsilon^{IJKL} \,Tr\Bigg[h^{k,\rho}_{\alpha_{v1}}\,\tau^{k,\rho}_{IJ}\,\tau^{k,\rho}_{KL}\,h^{k,\rho}_{\alpha_{v2}}\Bigg]\,,
\end{equation} 
where we choose the three surfaces $S^a_x$ such that they intersect $\alpha$ in one and the same point $x=\alpha_{1}\cap \alpha_{2}$ (this is always possible by refining). We can give the following graphical description of the operator $\hat{C}_{ren}^{ab}(x)$

\vspace{-1cm}

\begin{picture}(160,55)(0,0)
\unitlength 0.5mm
\linethickness{0.2mm}
\put(85,50){\line(1,0){25}}
\linethickness{0.2mm}
\put(80,50){\line(1,0){5}}
\put(80,50){\vector(-1,0){0.12}}
\linethickness{0.2mm}
\put(50,50){\line(1,0){30}}
\put(80,45){\makebox(0,0)[cc]{$k,\rho$}}

\put(115,50){\makebox(0,0)[cc]{=}}

\put(32,50){\makebox(0,0)[cc]{$\hat{\mathcal{C}}^{ab}_{ren}(x)$}}

\put(150,50){\makebox(0,0)[cc]{$-o_{(\alpha,S_x^a)}\,o_{(\alpha,S_x^b)}$}}

\linethickness{0.2mm}
\put(215,50){\line(1,0){25}}
\linethickness{0.2mm}
\put(210,50){\line(1,0){5}}
\put(210,50){\vector(-1,0){0.12}}
\linethickness{0.2mm}
\put(180,50){\line(1,0){30}}
\put(210,45){\makebox(0,0)[cc]{$k,\rho$}}
\multiput(210,46)(0,1.82){6}{\line(0,1){0.91}}
\put(210,63){\makebox(0,0)[cc]{$\epsilon^{IJKL}\,\tau_{IJ}\,\tau_{KL}$}}

\end{picture}

where the dashed lines, denoting the insertion of Lorentz generators, are placed in $x$. 
It is worth noting that the operator acting on the right-hand side of the equation above coincide with one of the Casimir operators of the Lorentz group, namely
\be
\hat{C}_2=\epsilon^{IJKL}\,\tau^{k,\rho}_{IJ}\,\tau^{k,\rho}_{KL}= 2k\rho\,,\label{C2}
\ee
such that we get
\be
\hat{C}^{ab}_{ren}(x)\,W^{k,\rho}_\alpha(\omega)=o_{(\alpha,S_x^a)}\,o_{(\alpha,S_x^b)}\,\,2k\rho\,W^{k,\rho}_\alpha(\omega)\,,
\ee
which coincides with the expression in \eqref{11}.

\section{Action of the scalar constraint}\label{B}

Let us compute the action of the scalar constraint operator \eqref{Sv} on the Wilson loop and, in order to have a nontrivial action, we choose one of the links of $\Omega^x_{ab}$ to coincide with a link of $\alpha$ beginning in $x$, the other link being orthogonal (by explicit computation one can verify that in all other cases $\mathcal{S}$ has a vanishing action).

The action of the momenta just provides the insertion of Lorentz generators at dual links, while that of the holonomy adds new links to $\alpha$ and provides the product between the holonomies in the representations $\{0,\rho_0\}$ and $\{0,\rho\}$ for the shared link, which can be computed via recoupling theory. The second term in the sum \eqref{Sv} vanishes, since under the exchange $I\leftrightarrow J$ the action of the two momenta provides a symmetric expression, while the operator $Tr\left[\tau^{0,\rho_0}_{IJ}\,h^{0,\rho_0}_{\Omega^v_{ab}}\right]$ is skew-symmetric. Hence, let us compute the action of the first term of the sum \eqref{Sv}: 

\begin{picture}(240,80)(0,0)
\unitlength 0.7mm
\linethickness{0.2mm}
\put(20,50){\line(1,0){25}}
\linethickness{0.2mm}
\put(20,50){\line(1,0){5}}
\put(20,50){\vector(-1,0){0.12}}
\linethickness{0.2mm}
\put(-5,50){\line(1,0){30}}
\put(20,45){\makebox(0,0)[cc]{$\rho$}}

\linethickness{0.2mm}
\put(105,50){\line(1,0){10}}
\linethickness{0.2mm}
\put(115,50){\line(1,0){10}}
\put(115,50){\vector(-1,0){0.12}}
\linethickness{0.2mm}
\put(95,50){\line(0,1){20}}
\linethickness{0.2mm}
\put(105,80){\line(1,0){10}}
\put(115,80){\vector(-1,0){0.12}}
\linethickness{0.2mm}
\put(115,80){\line(1,0){10}}
\linethickness{0.2mm}
\put(135,50){\line(0,1){20}}
\linethickness{0.2mm}
\linethickness{0.2mm}
\linethickness{0.2mm}
\put(140,50){\line(1,0){10}}
\put(140,50){\vector(-1,0){0.12}}
\linethickness{0.2mm}
\put(125,50){\line(1,0){15}}
\linethickness{0.2mm}
\linethickness{0.2mm}
\put(90,50){\line(1,0){15}}
\put(90,50){\vector(-1,0){0.12}}
\linethickness{0.2mm}
\multiput(125,45)(0,1.82){6}{\line(0,1){0.91}}
\linethickness{0.2mm}
\multiput(130,60)(1.82,0){6}{\line(1,0){0.91}}
\linethickness{0.2mm}
\linethickness{0.2mm}
\put(85,50){\line(1,0){5}}
\linethickness{0.2mm}
\put(95,70){\line(0,1){10}}
\linethickness{0.2mm}
\put(95,80){\line(1,0){10}}
\linethickness{0.2mm}
\put(125,80){\line(1,0){10}}
\linethickness{0.2mm}
\put(135,70){\line(0,1){10}}
\put(90,45){\makebox(0,0)[cc]{$\rho$}}

\put(145,45){\makebox(0,0)[cc]{$\rho$}}

\put(115,75){\makebox(0,0)[cc]{$\rho_0$}}

\put(115,45){\makebox(0,0)[cc]{$k',\rho'$}}

\put(150,60){\makebox(0,0)[cc]{$\tau_{IJ}\tau_{K}^{\phantom1J}$}}

\put(125,40){\makebox(0,0)[cc]{$\tau^{IK}$}}

\linethickness{0.2mm}
\put(170,50){\line(1,0){5}}
\linethickness{0.2mm}
\put(175,50){\line(1,0){25}}
\put(175,50){\vector(-1,0){0.12}}
\linethickness{0.2mm}
\put(200,50){\line(1,0){10}}
\put(200,50){\vector(-1,0){0.12}}
\linethickness{0.2mm}
\put(210,50){\line(1,0){25}}
\linethickness{0.2mm}
\put(235,50){\line(1,0){5}}
\put(235,50){\vector(-1,0){0.12}}
\linethickness{0.2mm}
\multiput(230,45)(0,1.82){6}{\line(0,1){0.91}}
\put(230,40){\makebox(0,0)[cc]{$\tau^{IK}$}}

\linethickness{0.2mm}
\put(220,50){\line(0,1){20}}
\linethickness{0.2mm}
\put(220,70){\line(0,1){10}}
\linethickness{0.2mm}
\put(210,80){\line(1,0){10}}
\linethickness{0.2mm}
\put(200,80){\line(1,0){10}}
\linethickness{0.2mm}
\put(190,80){\line(1,0){10}}
\put(200,80){\vector(-1,0){0.12}}
\linethickness{0.2mm}
\put(180,80){\line(1,0){10}}
\linethickness{0.2mm}
\put(180,70){\line(0,1){10}}
\linethickness{0.2mm}
\put(180,50){\line(0,1){20}}
\put(200,75){\makebox(0,0)[cc]{$\rho_0$}}

\linethickness{0.2mm}
\multiput(215,60)(1.82,0){6}{\line(1,0){0.91}}
\put(235,60){\makebox(0,0)[cc]{$\tau_{IJ}\tau_{K}^{\phantom1J}$}}

\put(175,45){\makebox(0,0)[cc]{$\rho$}}

\put(200,45){\makebox(0,0)[cc]{$k',\rho'$}}

\put(240,45){\makebox(0,0)[cc]{$\rho$}}

\put(65,50){\makebox(0,0)[cc]{$=\sum_{k'}\int d\mu(k',\rho')$}}

\put(-15,50){\makebox(0,0)[cc]{$\hat{\mathcal{S}}_{ren}(x)$}}

\put(160,50){\makebox(0,0)[cc]{$+$}}


\end{picture}

where the trivalent nodes represent Clebsch-Gordan coefficients and the insertion of Lorentz generators occurs in the (left, right and up) neighborhoods of the nearest node. The integral extends over all the admissible values of $\rho'$, {\it i.e.} those for which the Clebsch-Gordan coefficients at nodes do not vanish, and the measure $\mu(k',\rho')$ is proportional to $k^2+\rho^2$. Let us note that we can replace $\tau_{[I|J|}\tau_{K]}^{\phantom1J}$ with $\tau_{IK}$. It is worth noting the role of the arrows, which tell us about the orientation of each link. Using the invariance under Lorentz transformations, one can move the generators across the nodes using the following relations

\begin{picture}(135,90)(0,0)
\unitlength 0.7mm
\linethickness{0.2mm}
\put(20,80){\line(0,1){10}}
\put(20,80){\vector(0,1){0.12}}
\linethickness{0.2mm}
\linethickness{0.2mm}
\put(30,70){\line(1,0){10}}
\put(30,70){\vector(-1,0){0.12}}
\linethickness{0.2mm}
\linethickness{0.2mm}
\put(10,70){\line(1,0){20}}
\put(10,70){\vector(-1,0){0.12}}
\linethickness{0.2mm}
\put(20,70){\line(0,1){10}}
\linethickness{0.2mm}
\linethickness{0.2mm}
\put(0,70){\line(1,0){10}}
\linethickness{0.2mm}
\put(80,80){\line(0,1){10}}
\put(80,80){\vector(0,1){0.12}}
\linethickness{0.2mm}
\linethickness{0.2mm}
\put(90,70){\line(1,0){10}}
\put(90,70){\vector(-1,0){0.12}}
\linethickness{0.2mm}
\put(70,70){\line(1,0){20}}
\put(70,70){\vector(-1,0){0.12}}
\linethickness{0.2mm}
\put(80,70){\line(0,1){10}}
\linethickness{0.2mm}
\put(60,70){\line(1,0){10}}
\linethickness{0.2mm}
\put(140,80){\line(0,1){10}}
\put(140,80){\vector(0,1){0.12}}
\linethickness{0.2mm}
\linethickness{0.2mm}
\put(150,70){\line(1,0){10}}
\put(150,70){\vector(-1,0){0.12}}
\linethickness{0.2mm}
\put(130,70){\line(1,0){20}}
\put(130,70){\vector(-1,0){0.12}}
\linethickness{0.2mm}
\put(140,70){\line(0,1){10}}
\linethickness{0.2mm}
\put(120,70){\line(1,0){10}}
\put(50,70){\makebox(0,0)[cc]{=}}

\put(110,70){\makebox(0,0)[cc]{+}}

\linethickness{0.2mm}
\multiput(15,65)(0,1.82){6}{\line(0,1){0.91}}
\linethickness{0.2mm}
\multiput(85,65)(0,1.82){6}{\line(0,1){0.91}}
\linethickness{0.2mm}
\linethickness{0.2mm}
\multiput(135,75)(1.82,0){6}{\line(1,0){0.91}}
\put(15,60){\makebox(0,0)[cc]{$\tau_{IJ}$}}

\put(85,60){\makebox(0,0)[cc]{$\tau_{IJ}$}}

\put(150,75){\makebox(0,0)[cc]{$\tau_{IJ}$}}

\end{picture}

\vspace{-2cm}

so getting

\begin{picture}(240,80)(0,0)
\unitlength 0.7mm
\linethickness{0.2mm}
\put(20,50){\line(1,0){25}}
\linethickness{0.2mm}
\put(20,50){\line(1,0){5}}
\put(20,50){\vector(-1,0){0.12}}
\linethickness{0.2mm}
\put(-5,50){\line(1,0){30}}
\put(20,45){\makebox(0,0)[cc]{$\rho$}}

\linethickness{0.2mm}
\put(105,50){\line(1,0){10}}
\linethickness{0.2mm}
\put(115,50){\line(1,0){10}}
\put(115,50){\vector(-1,0){0.12}}
\linethickness{0.2mm}
\put(95,50){\line(0,1){20}}
\linethickness{0.2mm}
\put(105,80){\line(1,0){10}}
\put(115,80){\vector(-1,0){0.12}}
\linethickness{0.2mm}
\put(115,80){\line(1,0){10}}
\linethickness{0.2mm}
\put(135,50){\line(0,1){20}}
\linethickness{0.2mm}
\linethickness{0.2mm}
\linethickness{0.2mm}
\put(140,50){\line(1,0){10}}
\put(140,50){\vector(-1,0){0.12}}
\linethickness{0.2mm}
\put(125,50){\line(1,0){15}}
\linethickness{0.2mm}
\linethickness{0.2mm}
\put(90,50){\line(1,0){15}}
\put(90,50){\vector(-1,0){0.12}}
\linethickness{0.2mm}
\multiput(125,45)(0,1.82){6}{\line(0,1){0.91}}
\linethickness{0.2mm}
\linethickness{0.2mm}
\linethickness{0.2mm}
\put(85,50){\line(1,0){5}}
\linethickness{0.2mm}
\put(95,70){\line(0,1){10}}
\linethickness{0.2mm}
\put(95,80){\line(1,0){10}}
\linethickness{0.2mm}
\put(125,80){\line(1,0){10}}
\linethickness{0.2mm}
\put(135,70){\line(0,1){10}}
\put(90,45){\makebox(0,0)[cc]{$\rho$}}

\put(145,45){\makebox(0,0)[cc]{$\rho$}}

\put(115,75){\makebox(0,0)[cc]{$\rho_0$}}

\put(115,45){\makebox(0,0)[cc]{$k',\rho'$}}


\put(125,40){\makebox(0,0)[cc]{$\tau^{IK}\,\tau_{IK}$}}

\linethickness{0.2mm}
\put(170,50){\line(1,0){5}}
\linethickness{0.2mm}
\put(175,50){\line(1,0){25}}
\put(175,50){\vector(-1,0){0.12}}
\linethickness{0.2mm}
\put(200,50){\line(1,0){10}}
\put(200,50){\vector(-1,0){0.12}}
\linethickness{0.2mm}
\put(210,50){\line(1,0){25}}
\linethickness{0.2mm}
\put(235,50){\line(1,0){5}}
\put(235,50){\vector(-1,0){0.12}}
\linethickness{0.2mm}
\multiput(230,45)(0,1.82){6}{\line(0,1){0.91}}
\put(230,40){\makebox(0,0)[cc]{$\tau_{IK}\,\tau^{IK}$}}

\linethickness{0.2mm}
\put(220,50){\line(0,1){20}}
\linethickness{0.2mm}
\put(220,70){\line(0,1){10}}
\linethickness{0.2mm}
\put(210,80){\line(1,0){10}}
\linethickness{0.2mm}
\put(200,80){\line(1,0){10}}
\linethickness{0.2mm}
\put(190,80){\line(1,0){10}}
\put(200,80){\vector(-1,0){0.12}}
\linethickness{0.2mm}
\put(180,80){\line(1,0){10}}
\linethickness{0.2mm}
\put(180,70){\line(0,1){10}}
\linethickness{0.2mm}
\put(180,50){\line(0,1){20}}
\put(200,75){\makebox(0,0)[cc]{$\rho_0$}}

\linethickness{0.2mm}

\put(175,45){\makebox(0,0)[cc]{$\rho$}}

\put(200,45){\makebox(0,0)[cc]{$k',\rho'$}}

\put(240,45){\makebox(0,0)[cc]{$\rho$}}

\put(65,50){\makebox(0,0)[cc]{$=\sum_{k'}\int d\mu(k',\rho')$}}

\put(-15,50){\makebox(0,0)[cc]{$\hat{\mathcal{S}}_{ren}(x)$}}

\put(160,50){\makebox(0,0)[cc]{$+$}}

\end{picture}

\vspace{-4cm}

\begin{picture}(195,80)(0,0)
\unitlength 0.7mm
\linethickness{0.2mm}
\put(150,50){\line(1,0){10}}
\linethickness{0.2mm}
\put(160,50){\line(1,0){10}}
\put(160,50){\vector(-1,0){0.12}}
\linethickness{0.2mm}
\put(140,50){\line(0,1){20}}
\linethickness{0.2mm}
\put(150,80){\line(1,0){10}}
\put(160,80){\vector(-1,0){0.12}}
\linethickness{0.2mm}
\put(160,80){\line(1,0){10}}
\linethickness{0.2mm}
\put(180,50){\line(0,1){20}}
\linethickness{0.2mm}
\linethickness{0.2mm}
\linethickness{0.2mm}
\put(185,50){\line(1,0){10}}
\put(185,50){\vector(-1,0){0.12}}
\linethickness{0.2mm}
\put(170,50){\line(1,0){15}}
\linethickness{0.2mm}
\linethickness{0.2mm}
\put(135,50){\line(1,0){15}}
\put(135,50){\vector(-1,0){0.12}}
\linethickness{0.2mm}
\linethickness{0.2mm}
\put(130,50){\line(1,0){5}}
\linethickness{0.2mm}
\put(140,70){\line(0,1){10}}
\linethickness{0.2mm}
\put(140,80){\line(1,0){10}}
\linethickness{0.2mm}
\put(170,80){\line(1,0){10}}
\linethickness{0.2mm}
\put(180,70){\line(0,1){10}}
\put(135,45){\makebox(0,0)[cc]{$\rho$}}

\put(190,45){\makebox(0,0)[cc]{$\rho$}}

\put(160,75){\makebox(0,0)[cc]{$\rho_0$}}

\put(160,45){\makebox(0,0)[cc]{$k',\rho'$}}

\put(75,50){\makebox(0,0)[cc]{$=\sum_{k'}\int d\mu(k',\rho')\, \left[C_1(k',\rho')-C_1(0,\rho)\right]$}}

\end{picture}

where the other Casimir operator of the Lorentz group appears, 
\be
\hat{C}_1=\tau^{k,\rho,IJ}\,\tau^{k,\rho}_{IJ}=k^2-\rho^2-1\,.\label{C1}
\ee
In the limit $\epsilon\rightarrow 0$, the holonomies along the loop go to the identity and one ends up with only the two Clebsch-Gordan coefficients at nodes, {\it i.e.}

\begin{picture}(165,80)(0,0)
\unitlength 0.7mm
\linethickness{0.2mm}
\put(10,50){\line(1,0){10}}
\linethickness{0.2mm}
\put(20,50){\line(1,0){10}}
\put(20,50){\vector(-1,0){0.12}}
\linethickness{0.2mm}
\put(0,50){\line(0,1){20}}
\linethickness{0.2mm}
\put(10,80){\line(1,0){10}}
\put(20,80){\vector(-1,0){0.12}}
\linethickness{0.2mm}
\put(20,80){\line(1,0){10}}
\linethickness{0.2mm}
\put(40,50){\line(0,1){20}}
\linethickness{0.2mm}
\linethickness{0.2mm}
\linethickness{0.2mm}
\put(45,50){\line(1,0){10}}
\put(45,50){\vector(-1,0){0.12}}
\linethickness{0.2mm}
\put(30,50){\line(1,0){15}}
\linethickness{0.2mm}
\linethickness{0.2mm}
\put(-5,50){\line(1,0){15}}
\put(-5,50){\vector(-1,0){0.12}}
\linethickness{0.2mm}
\linethickness{0.2mm}
\put(-10,50){\line(1,0){5}}
\linethickness{0.2mm}
\put(0,70){\line(0,1){10}}
\linethickness{0.2mm}
\put(0,80){\line(1,0){10}}
\linethickness{0.2mm}
\put(30,80){\line(1,0){10}}
\linethickness{0.2mm}
\put(40,70){\line(0,1){10}}
\put(-5,45){\makebox(0,0)[cc]{$\rho$}}

\put(50,45){\makebox(0,0)[cc]{$\rho$}}

\put(20,75){\makebox(0,0)[cc]{$\rho_0$}}

\put(20,45){\makebox(0,0)[cc]{$k',\rho'$}}

\put(60,50){\makebox(0,0)[cc]{$\rightarrow$}}

\linethickness{0.2mm}
\put(70,50){\line(1,0){5}}
\put(70,50){\vector(-1,0){0.12}}
\linethickness{0.2mm}
\put(65,50){\line(1,0){5}}
\put(70,45){\makebox(0,0)[cc]{$\rho$}}

\put(135,50){\makebox(0,0)[cc]{$\displaystyle\sum_{\substack{j',j_0\\m',m_0}}\left(C^{0\rho j_i m_i}_{k'\rho' j' m'\,0\rho_0j_0m_0}\right)^*\, C^{0\rho j_f m_f}_{k'\rho' j' m'\,0\rho_0j_0m_0}$}}

\linethickness{0.2mm}
\put(195,50){\line(1,0){10}}
\put(195,50){\vector(-1,0){0.12}}
\linethickness{0.2mm}
\put(190,50){\line(1,0){5}}
\put(200,45){\makebox(0,0)[cc]{$\rho$}}

\end{picture}

$j_i$, $m_i$ and $j_f$, $m_f$ being spin numbers which are contracted with those of the right and left holonomies in the representation $\rho$. Hence, in order to find physical states, we must solve the following condition for $\rho$
\begin{align}
\sum_{k'}\int d\mu(k',\rho')\,(-k'^2+\rho')^2\,\displaystyle\sum_{\substack{j',j_0\\m',m_0}}\, &\left(C^{0\rho j_i m_i}_{k'\rho' j' m'\,0\rho_0j_0m_0}\right)^*\, C^{0\rho j_f m_f}_{k'\rho' j' m'\,0\rho_0j_0m_0} = \nonumber\\
&\rho^2\,\sum_{k'}\int d\mu(k',\rho')\,\displaystyle\sum_{\substack{j',j_0\\m',m_0}}\, \left(C^{0\rho j_i m_i}_{k'\rho' j' m'\,0\rho_0j_0m_0}\right)^*\, C^{0\rho j_f m_f}_{k'\rho' j' m'\,0\rho_0j_0m_0}\,.\label{fc}
\end{align}
It is worth noting that 
\be
\displaystyle\sum_{\substack{j',j_0\\m',m_0}}\left(C^{k_1\rho_1 j_i m_i}_{k'\rho' j' m'\,0\rho_0j_0m_0}\right)^*\, C^{k_2\rho_2 j_f m_f}_{k'\rho' j' m'\,0\rho_0j_0m_0}
=\delta(\rho_1-\rho_2) \,\delta_{k_1,k_2}\, \delta_{j_i,j_f}\,\delta_{m_i,m_f}\,,
\ee
which implies that the final result is the original Wilson loop times some factor depending on the summation of two Clebsch-Gordon coefficients. Therefore, in the continuum limit, the action of the scalar constraint does not take Wilson loops out of the kernel of the operator $\hat{C}^{ab}$ and the restriction to $k=0$ is consistent with the dynamics. This formally solves the issues related with the second-class character of the system of constraints \eqref{hcon}.

\end{document}